\documentclass[prl,aps,floats,amsmath,amssymb,twocolumn,preprintnumbers,nofootinbib]{revtex4}
\usepackage{textcomp}
\usepackage{amsmath}
\usepackage{amssymb}
\usepackage{epsfig}
\usepackage{tabularx}
\usepackage{graphicx}
\usepackage{inputenc}
\usepackage{url}
\begin{document}
\title{Higgs finder and mass estimator}
\author{Vernon Barger, Peisi Huang}
\affiliation{Department of Physics, University of Wisconsin, Madison, WI 53706, USA}
\begin{abstract}
We exploit the spin and kinematic correlations in the decay of a scalar boson into a pair of real or virtual W-bosons, with both W-bosons decaying leptonically,  for Higgs boson discovery at 7 TeV LHC energy with 10 $fb^{-1}$ luminosity. Without reconstruction of the events, we obtain estimators of the higgs mass from the peak and width of the signal distribution in $m_{ll}$. The separation of  signal and background with other distributions, such as the azimuthal angle between two W decay planes, the rapidity difference between the two leptons, $\not{\!{\rm E}}_T$ and the $p_T$ of leptons, are also prescribed.  Our approach identifies  the salient higgs to dilepton signatures that allow subtraction of the continuum W*W* background.
\end{abstract}
\maketitle
\begin{figure}[here]
\includegraphics[width = 0.45 \textwidth]{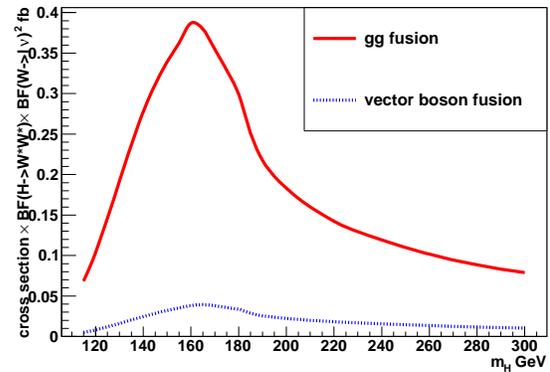}
\caption{SM Higgs production cross section times the branching fractions to two Ws that decay leptonically. $l = e, \mu$}
\label{fig:xsec}
\end{figure}

The higgs boson is the only missing brick of the Standard Model (SM) \cite{Djouadi:2005gi}. The $h \rightarrow W^{+} W^{-} \rightarrow l \nu l \nu$ channel has been of long interest for higgs discovery\cite{Keung:1984hn}\cite{Glover:1988fn}\cite{Barr:2009mx}\cite{Barger:1990mn}\cite{Barger:1993qu}\cite{Han:1998sp}, because of its relatively clean signal and the large branching fraction for $m_h$ near $2m_W$.  The CDF and DO experiments at the Tevatron  and the ATLAS and CMS experiments at the LHC have searched for the $h \rightarrow W^{*}W^{*} \rightarrow \mu \overline{\nu}_{\mu} \nu_{\mu}\overline{\mu}$ process and have excluded a SM higgs in a range of $m_{h}$ around 166 GeV \cite{Aaltonen:2010yv}\cite{Aaltonen:2010cm}\cite{Abazov:2011bc}\cite{Aad:2011qi}\cite{Chatrchyan:2011tz}. The SM higgs production cross section times the branching fraction to two Ws in the SM is plotted in figure \ref{fig:xsec}. The maximum $h\rightarrow W^{*}W^{*}$ signal from gluon fusion is at $m_h$ = 165 GeV. The dominant production at $m_h < 1$ TeV occurs via the parton subprocess $ gluon + gluon \rightarrow h$ and WW-fusion takes over at  $m_h > $ 1 TeV\cite{Dittmaier:2011ti}.  Higgs production via gluon fusion could be larger than this estimate if extra colored states contribute to the gluon fusion loop \cite{PhysRevD.83.094018} or it could be smaller if the weak coupling is shared by two neutral Higgs states, as would be the case in supersymmetry \cite{Harlander:2003xy}, or if the higgs has invisible decay modes.

\begin{figure}[here]
\includegraphics[width = 0.45 \textwidth]{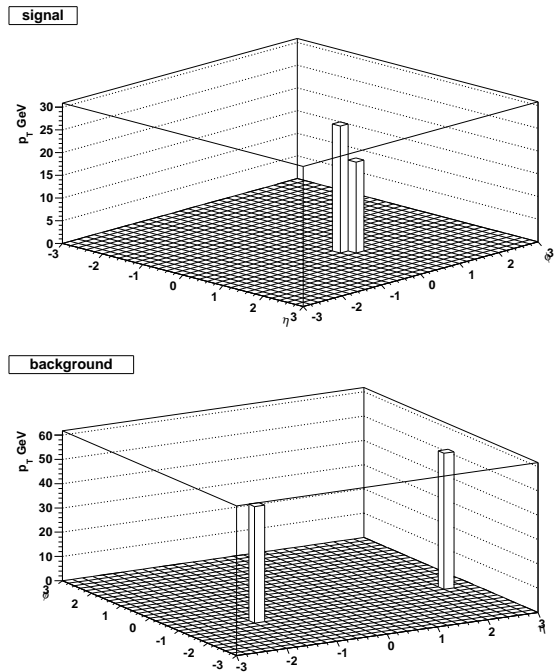}
\caption{Sample events for the $m_h$ = 160 GeV signal and the W*W* background with no jets.}
\label{fig:event}
\end{figure}

Many phenomenological studies have been made of the  $h \rightarrow W^*W^*$ signal\cite{Han:1998sp}\cite{Han:1998ma}\cite{Berger:2010nc}\cite{Anastasiou:2009bt}\cite{PhysRevD.44.2701} and that of the closely related $h \rightarrow Z^*Z^*$ channel \cite{Duncan1986393}\cite{Zecher:1994kb}\cite{Buszello:2002uu}\cite{Choi:2009hn}\cite{Dobrescu:2009zf}\cite{Gao:2010qx}.  The W*W* signal identification with leptonic W* decay is challenging. With two missing neutrinos, the events are not fully reconstructible. Also, the W*W* signal may have similar kinematics as the background. Since the background is much larger than the signal at the LHC, differences in the distributions of the signal and background must be used to identify and quantify the Higgs signal.  A typical signal event in this channel for $m_h = 160$ GeV is shown in the N(number of events) vs $\eta$ (rapidity difference of the leptons) vs $\phi$ (azimuthal angular difference of the leptons) plot in figure \ref{fig:event},  along with that of a sample background event, illustrating that there can be distinguishing features.  Our aim is to utilize the differences in the signal and background characteristics to enable a background subtraction and make a clear identification of any higgs signal, in novel ways that have not been fully explored in other studies.  Our approach relies on the SM prediction of the background distributions from the $q \overline{q} \rightarrow W^{*}W ^{*}$ subprocess at NLO order with the rejection of QCD jets.  The theory normalization of this background can be tested in ranges of the distributions where the higgs signal of a given $m_H$ does not contribute. Also, WZ production can serve as an independent calibration of the WW background, since the WZ final state does not have a neutral higgs signal contribution.

Nelson \cite{Nelson:1986ki} investigated the correlation between the two W decay planes, and used that to distinguish the higgs signal from the WW background. Choi et al\cite{Choi:2009hn}\cite{Choi:2010dw} studied the signal distributions in transverse mass variables\cite{nla.cat-vn2582768}.  Dobrescu and Lykken \cite{Dobrescu:2009zf} computed the fully differential width for higgs decays to $l\nu$$jj$, and constructed distributions of $m_{l\nu}$, $m_{jj}$, polar ($\theta_{l}$) and azimuthal($\phi_l$) angles between the charged lepton in the $l\nu$ rest frame and the $W^{+}$ in the higgs rest frame, and $\theta_{j}$, the angle between $-(\overrightarrow{p_{l}}+\overrightarrow{p_{\nu}})$ and the fastest jet direction in the higgs rest frame.

\begin{figure}[here]
\includegraphics[width = 0.45 \textwidth]{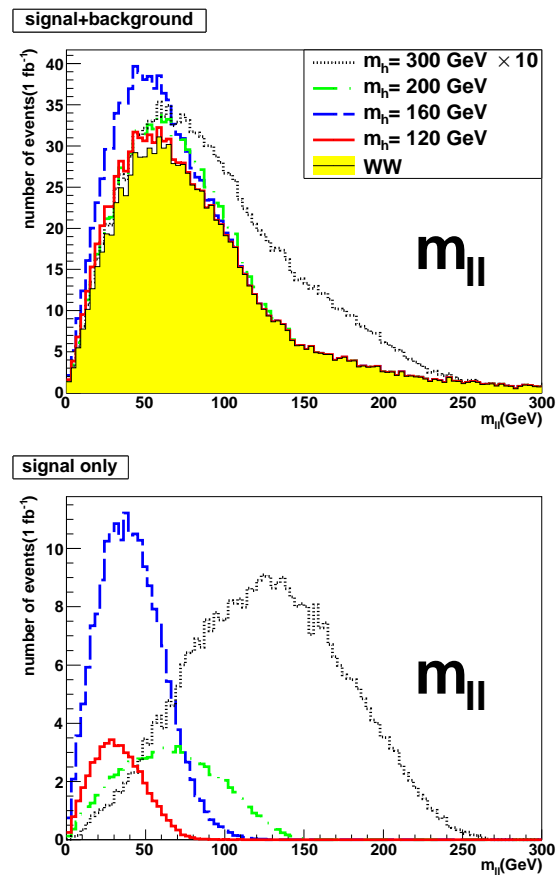}
\caption{$m_{ll}$ event distribution of the SM higgs signal at various $m_{h}$ and the background from continuum W*W* production for 1 $fb^{-1}$ luminosity at 7 TeV, summed over $l = e,\mu$}.
\label{fig:mll}
\end{figure}

\begin{figure}[here]
\includegraphics[width = 0.45 \textwidth]{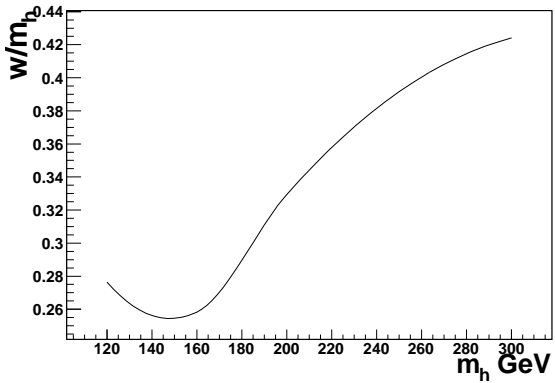}
\caption{Width (w) of the $m_{ll}$ distribution of the higgs signal compare to $m_{h}$. Note that at $m_h$ = 150 GeV(200 GeV),w $= \frac{1}{4}m_h$ (w $ = \frac{1}{3}m_h)$.}
\label{fig:width}
\end{figure}

\underline{Estimating$ $ the $ $higgs$ $ mass$ $ from$ $ the $ $invariant$ $ mass $ $of }
\underline{two $$leptons} \\
The matrix element for the higgs signal at is similar to that of muon decay, except for the placement of muon spinor and inclusion of off-shell W-propagators \cite{Nelson:1986ki}. 
We generated 200,000 events at four different higgs mass points and W*W* background with Sherpa \cite{Gleisberg:2008ta}, which includes the exact tree level matrix element and QCD radiation, at 7 TeV LHC center of mass (cm) energy. Jets are defined using the anti-kt algorithm \cite{Cacciari:2008gp} with R = 0.4 and the jet clusterings are implemented using the fastjet package\cite{Cacciari:2005hq}.  We use HiggsDecay \cite{Djouadi:1997yw} for calculation of the Higgs total and partial widths. We normalize the dilepton signal rate, $l = e, \mu$, for no jets to the NNLO  calculation \cite{Baglio:2010ae}, which is 104 $\mathrm{fb}$ at $m_{H} = $ 120 GeV, 389 $\mathrm{fb}$ at $m_{H} = $  160 GeV, 182 $\mathrm{fb}$ at $m_{H} = $ 200 GeV, and 83 $\mathrm{fb}$ at $m_{H} = $ 300 GeV. The $WW\rightarrow l \nu l \nu$ background is normalized to the NLO prediction \cite{Yang:1328692} of 2095 $\mathrm{fb}$. These cross sections are for the dilepton final states with l $=e,\mu $, including the leptonic  branching fractions. The $m_{ll}$ distributions, with and without the WW background, are given in figure \ref{fig:mll}, each for 1 $\mathrm{fb ^{-1}}$ integrated luminosity. The width(w) of the $m_{ll}$ distribution is given in figure \ref{fig:width}. This width is large compared to the total decay width of higgs, making it sensitive only to the higgs mass.  Here we only require two leptons and no jets, with no additional cuts.

The following empirical relationship between $m_{H}$ and $m_{ll}$ of the signal is found, where  $``$peak'' is the maximum and $``$end'' is the end point of the $m_{ll}$ distribution.
\begin{equation}
\begin{aligned}
&m_{H} = 2 (m_{llpeak})+m_{W} \\
&m_{H} = m_{llend}+\frac{m_{W}}{2}
\end{aligned}
\label{eq:mll}
\end{equation}
This relationship holds for all the higgs mass points, including when one W is off-shell, near the 2$m_W$ threshold and well above the threshold.
The signal and W*W* background within windows around the peak values of $m_{ll}$ are listed in Table \ref{table:mll}. The estimated significances $\frac{S}{\sqrt{S+B}}$ before the W*W* continuum background subtraction and  $``$idealized''$\sqrt{S}$ significances after subtraction of this background are given.  A significance $> 5$ could be achievable from the $m_{ll}$ distribution alone, once the background subtraction has been made.

\begin{figure}[here]
\includegraphics[width = 0.45 \textwidth] {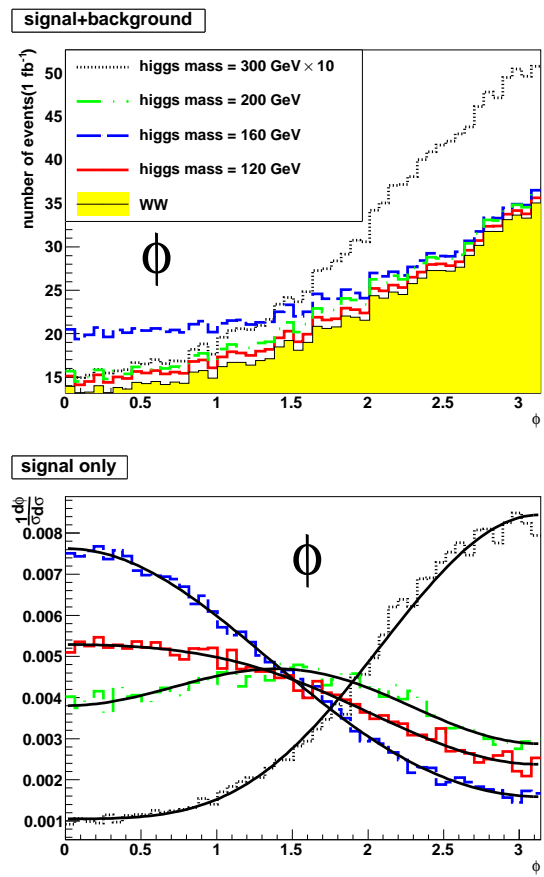}
\caption{The azimuthal angle between the two W decay planes.}
\label{fig:phi}
\end{figure}

\begin{figure}[here]
\includegraphics[width = 0.45 \textwidth] {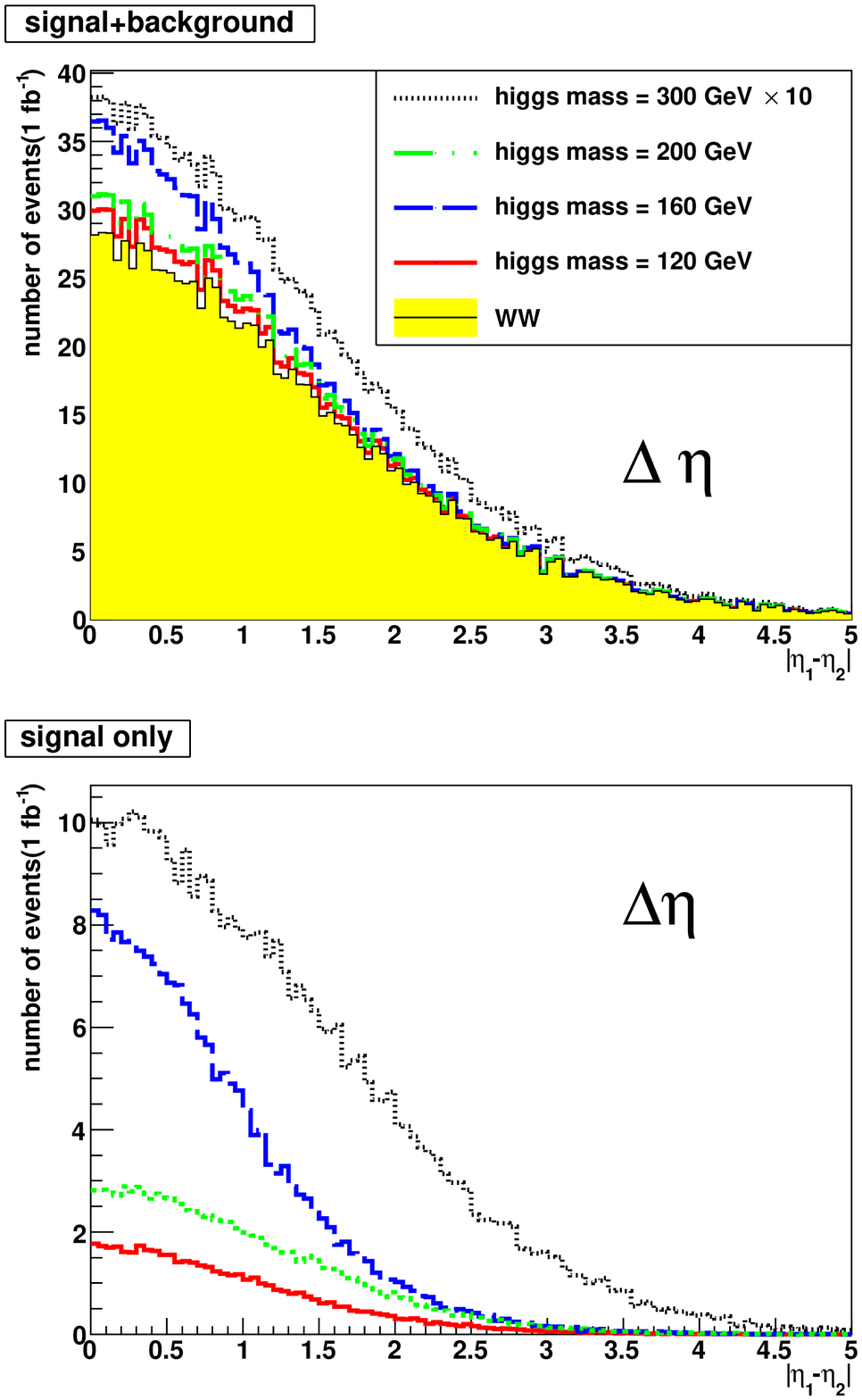}
\caption{Pseudorapidity difference $\Delta\eta = \mid\eta_1-\eta_2\mid$ of the two leptons}
\label{fig:eta}
\end{figure}

\begin{table}[htb]
\centering
\begin{tabular}{c|c|c|c|c|c|c}
\hline \hline
$m_h$ (GeV) & $m_{ll}$ window& signal & background& background & $\frac{S}{\sqrt{S+B}}$&$\sqrt{S}$   \\
&(GeV)& inside & inside & outside&& \\
&& window & window &window && \\
\hline
120 & $10 - 50$ & 373&2746&7723&2.1& 19\\
160 & $20 - 70$ & 1478 & 4326 &6144&6.1&38\\
200 & $30 - 110$ & 687 & 6713 &3756&2.5&26 \\
300 & $60 - 200$ & 324 & 5901 &4568 &1.3&18\\
\hline
\end{tabular}
\caption{The signal and background events at 7 TeV within the specified $m_{ll}$ windows around the peak values. The number of events in the signal and background columns are for 10 $\mathrm{fb^{-1}}$ integrated luminosity anticipated from ATLAS and CMS combined. Event numbers are summed over $l = e, \mu$.}
\label{table:mll}
\end{table}
\underline{Parametrization$ $ of$ $ the$ $ azimuthal $ $angular $ $ distribution}
The correlation function for the azimuthal angle between the two W decay planes can be parametrized as \cite{Nelson:1986ki}:
\begin{equation}
F(\phi) = 1 + \alpha cos\phi + \beta cos2\phi
\label{eq:phi}
\end{equation}
The direction of the normal to a W decay plane is defined as the cross product of momentum direction of the lepton with the beam direction. In figure \ref{fig:phi} we plot the $\phi$ distribution of signal and the WW background and fit the normalized distributions to Eq(\ref{eq:phi}). The resulting $\alpha$ and $\beta$ values are given in Table \ref{table:phi}.
\begin{table}[h]
\centering
\begin{tabular}{c|c|c|c|c|c}
\hline \hline
higgs mass (GeV) & 120 & 160 & 200 & 300& background \\
\hline
$\alpha$ & 0.36&0.68&0.12&-0.95&-0.43 \\
$\beta$&-0.06 &0.04&-0.17&0.22&0.09\\
\hline
\end{tabular}
\caption{The $\alpha$ and $\beta$ parametrization from fit of Eq (\ref{eq:phi}) to the $\phi$ distributions}
\label{table:phi}
\end{table}
\begin{figure*}[htb]
\includegraphics[width = 0.9 \textwidth] {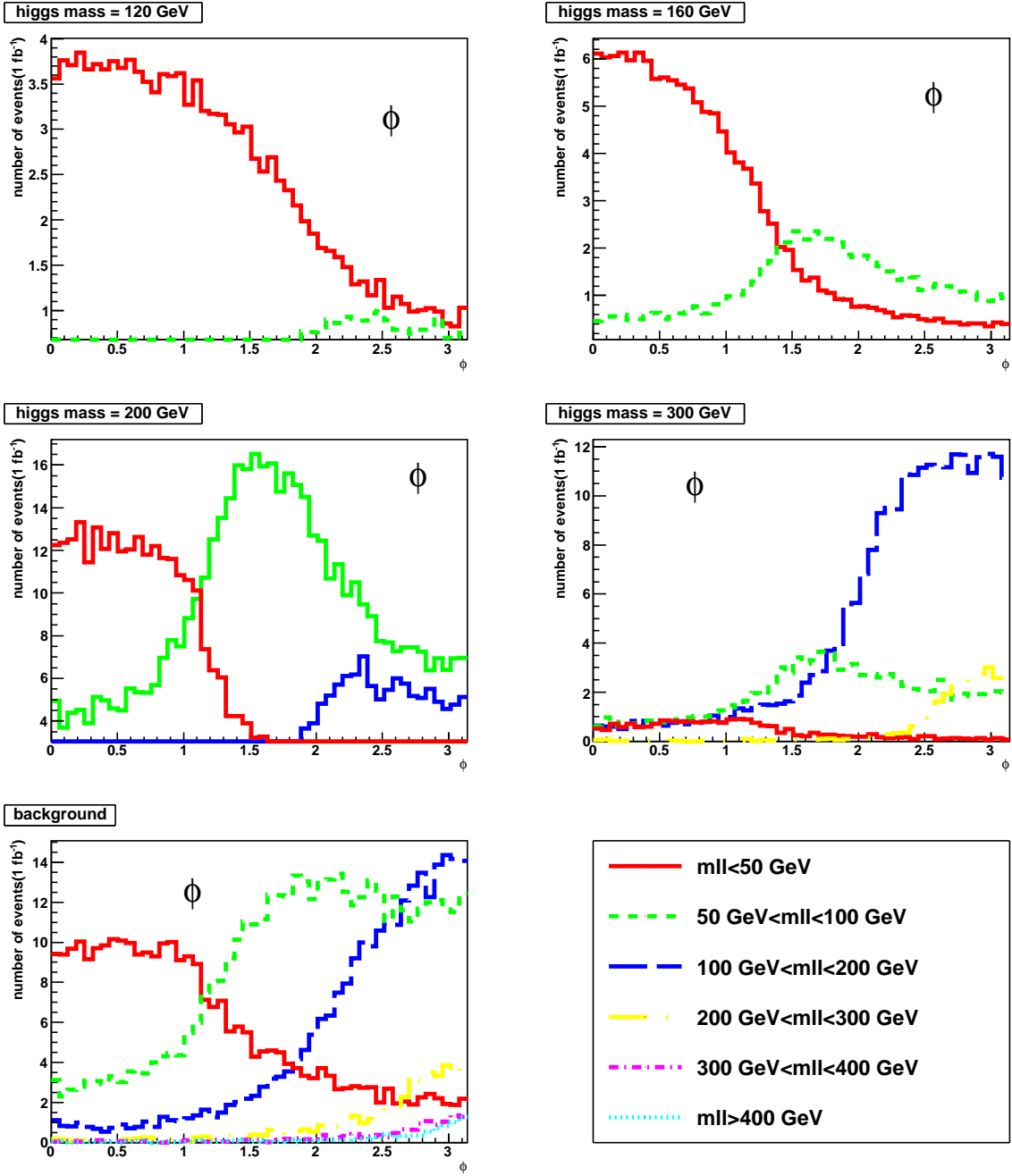}
\caption{$\phi$ distributions in different $m_{ll}$ bins of the higgs signals and the background.}
\label{fig:PhiVsMll}
\end{figure*}
It can be seen that  $\alpha > $ 0 in the transverse-transverse (TT) dominant region, while $\alpha <$ 0 in the longitudinal-longitudinal (LL) dominant region. At $m_H = 1 + \sqrt{17}  m_W = 182$ GeV,  $\Gamma(h \rightarrow W_{T} W_{T}) = \Gamma(h \rightarrow W_{L} W_{L})$.
The $\phi$ distribution at $m_H$ = 200 GeV is almost flat, as expected. The WW background has $\alpha <$ 0, because it is LL dominant. The $\phi$ distributions within different $m_{ll}$ bins are shown in Fig. \ref{fig:PhiVsMll}. In the $m_{ll} < 50$ bin, signal and background are both dominantly TT, and in the high $m_{ll}$ bin, both are dominantly LL.
The pseudorapidity difference $\Delta\eta = \mid\eta_1 - \eta_2\mid$ of the two leptons is plotted in Fig. \ref{fig:eta}. Note that the charged leptons from signal are closer in $\Delta\eta$ than for the background .

\begin{figure}[here]
\includegraphics[width = 0.45 \textwidth] {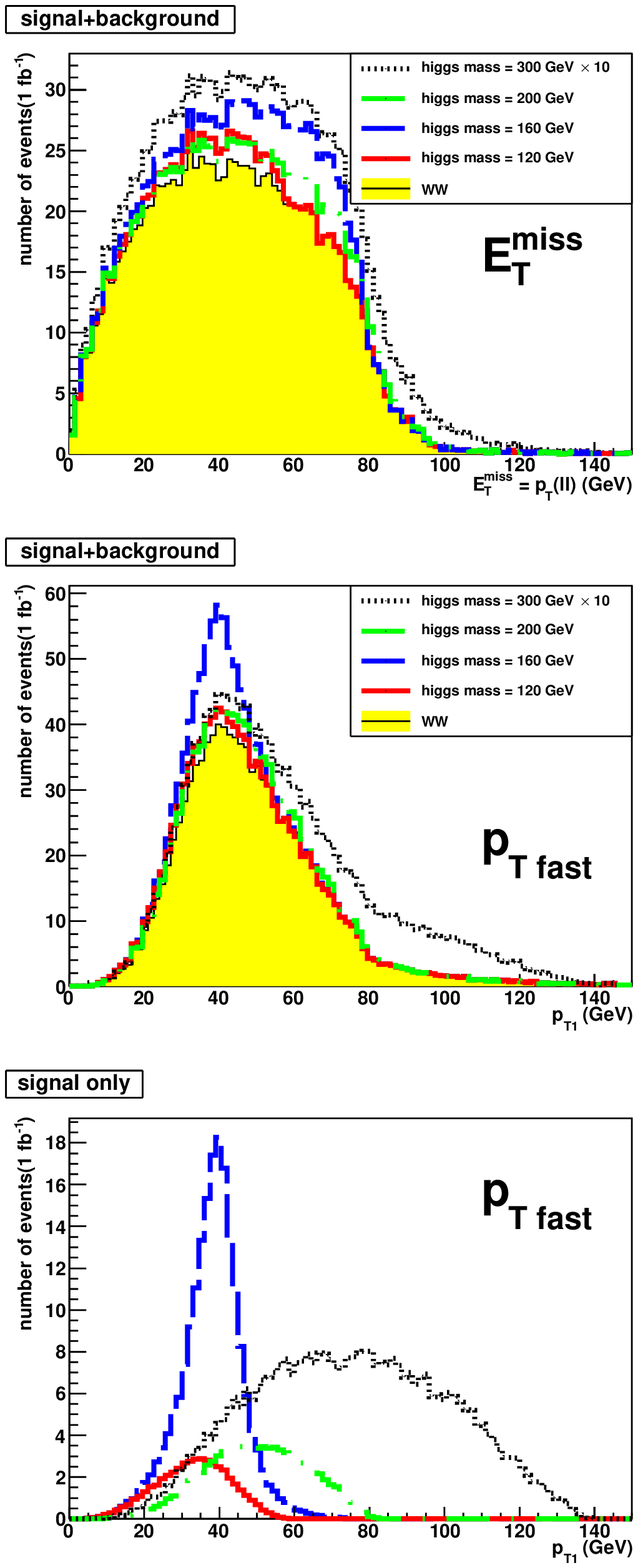}
\caption{The $\not{\!{\rm E}}_T=pt(ll)$ distribution and the $p_T$ distribution of the fastest lepton. Note the sharply peaked $p_{T1}$ from $m_H$ = 160 GeV.}
\label{fig:Etmiss}
\end{figure}

\begin{figure}[here]
\includegraphics[width = 0.45 \textwidth] {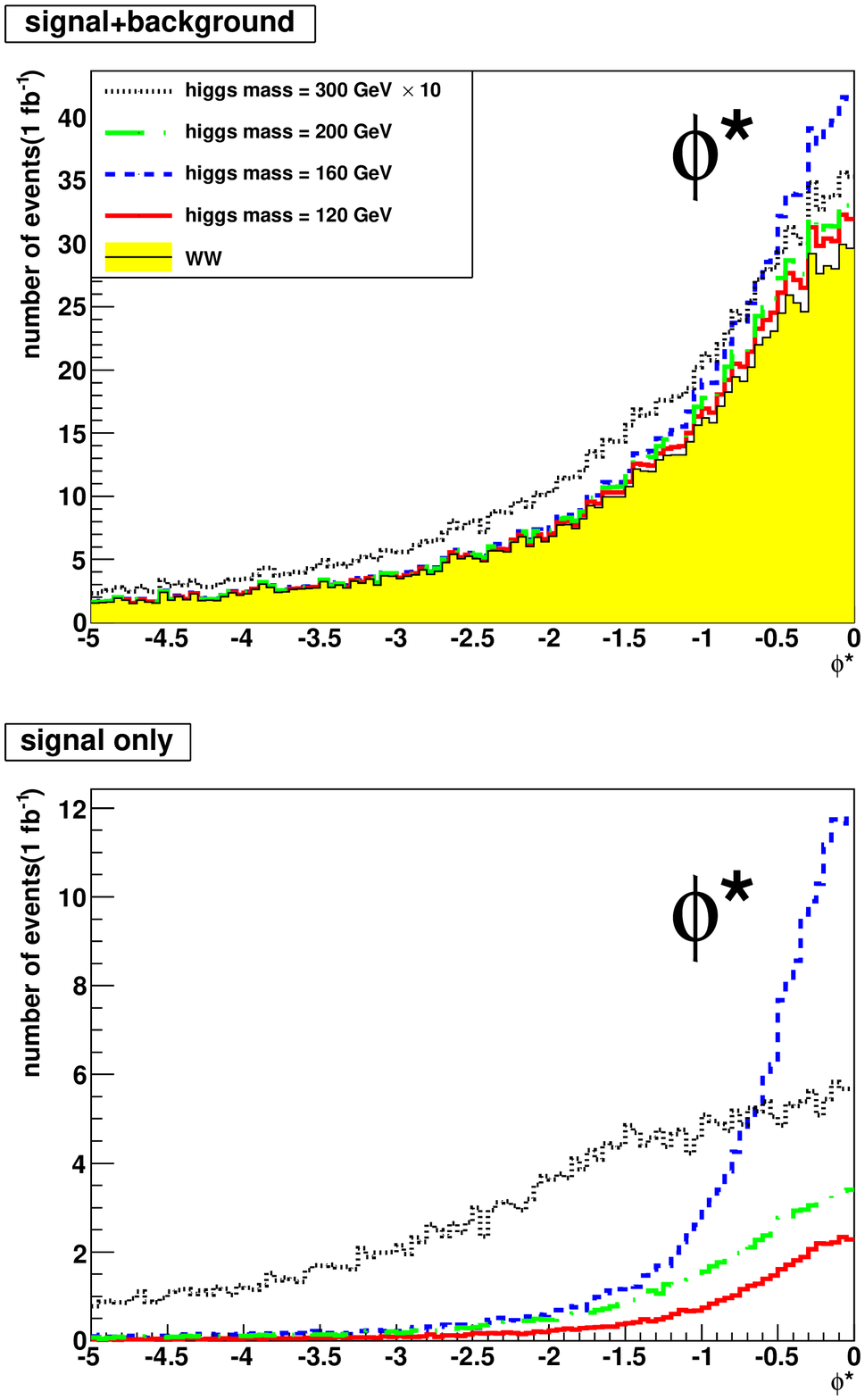}
\caption{$\phi$ * distribution.}
\label{fig:phistar}
\end{figure}

\underline{Background$  $ estimation}
Other variables can also differentiate signal from the background, such as $\not{\!{\rm E}}_T = p_{T}(ll)$ and the $p_T$ distribution of the fastest lepton, $p_{T1}$, shown in Fig. \ref{fig:Etmiss}.
The $p_T$ distribution of the fast lepton is very sensitive to the higgs mass. This distribution is sharply peaked for $m_h = 160$ GeV. A recent proposed variable, $\phi$* \cite{Banfi:2010cf} is plotted in Fig. \ref{fig:phistar}. $\phi$* is defined as $\phi$* = tan[($\pi$ - $\phi$)/2]sin$\theta$*, where $\phi$ is the azimuthal angle between the two leptons and cos$\theta$* = tanh[($\eta^{-}$-$\eta^+)$/2], with $\eta^-$ ($\eta^+$) being the pseudorapidity of the negatively charged lepton. It has been argued that $\phi$* may be more precisely determined than $\phi$.

\begin{figure}[here]
\includegraphics[width = 0.45 \textwidth] {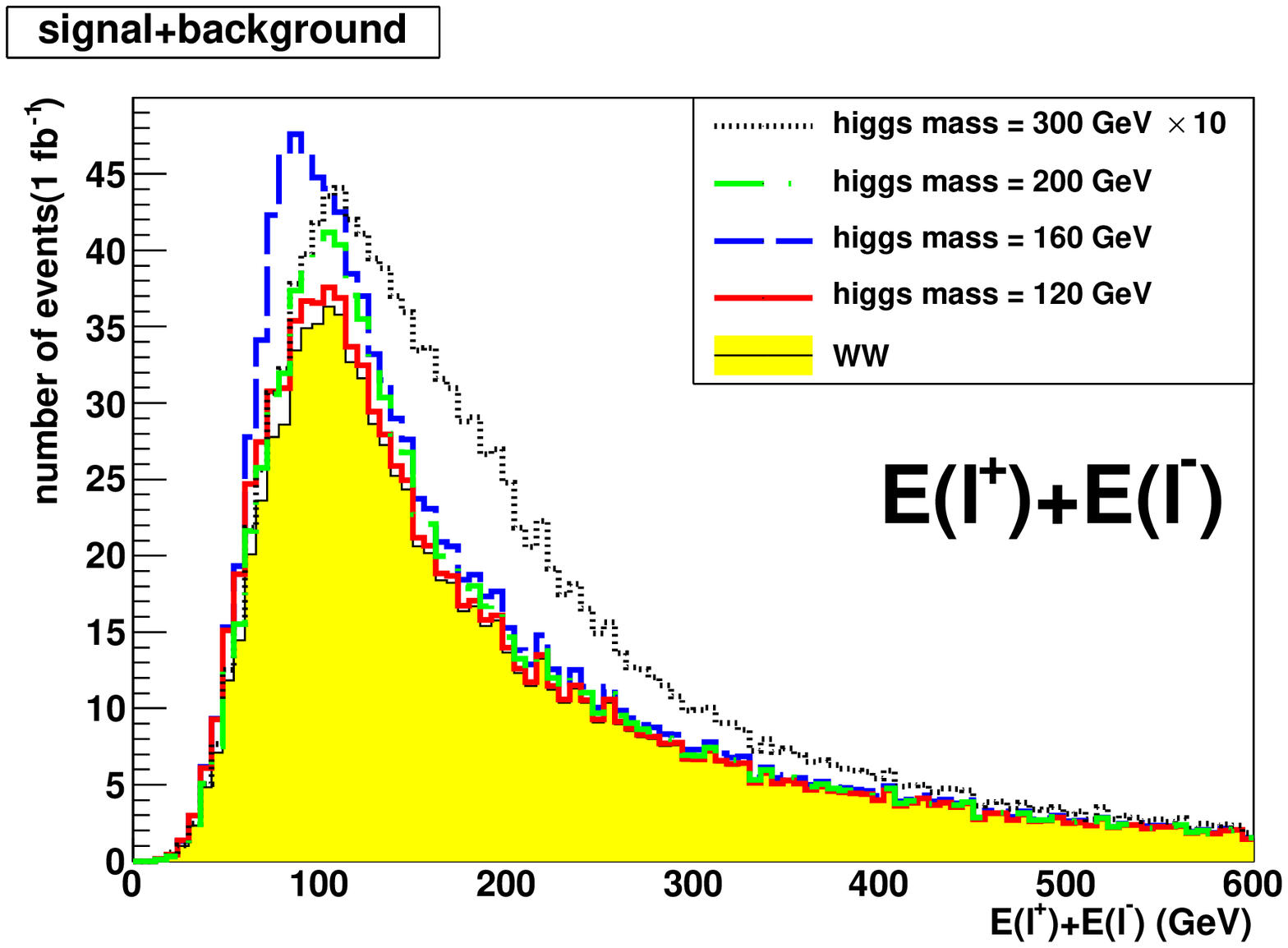}
\caption{Sum of energy of the two leptons.}
\label{fig:sum}
\end{figure}

The sum of the energy of the two leptons is shown in Fig. \ref{fig:sum}. The peak value of the $E(l^+)$ + $E(l^-)$ distribution of the signal is corelated with $m_H$.

Other backgrounds include $t\overline{t}$ pair production, single top production, W(or Z) + jets, Drell-Yan process (which does not contribute to the $e \mu$ events), and $\tau\overline{\tau}$ production. All these backgrounds can be suppressed by vetoing the jets and suitable cuts on the distributions of the variables discussed above. In the background subtraction, all of these backgrounds must be taken into account. The analysis of ATLAS shows that after reasonable cuts, all the other backgrounds are small compared to the W*W* background\cite{ATLAS-CONF-2011-005}.  Multivariable techniques, such as neural networks and boost decision trees, are another effective approach to background rejection. 

\underline{Conclusions$$ and$$ outlook} After subtracting the WW continuum background from the dilepton data, the higgs mass can be estimated using Eq (\ref{eq:mll}). The width of the $m_{ll}$ distribution provides another estimation of the higgs mass. The $\phi$, $\phi$* and $\eta$ distributions are almost unchanged by the experimental $p_{T}$ acceptance cuts. The $m_{ll}$, $p_T$ and E distributions are truncated at the lower ends by the $p_T$ and $\eta$ acceptance cuts. Our analysis techniques can be applied to scalars in other models that decay via the WW mode such as the radion\cite{Goldberger:1999un}\cite{Cheung:2000rw}\cite{Rizzo:2002pq}\cite{Kribs:2006mq}\cite{Davoudiasl:2010fb}\cite{Eshel:2011wz} or a dilaton \cite{Goldberger:2007zk}. The merit of the $m_{ll}$ peak estimator in Eq(\ref{eq:mll}) and width estimator in Fig.\ref{fig:width} is their simple dependences on the higgs boson mass.
\bibliographystyle{unsrt}
\bibliography{higgsref}
\end{document}